\documentclass[11pt]{elsart}
\usepackage{amsfonts}
\usepackage{slashbox}
\usepackage{graphicx}
\usepackage{amssymb,amsmath}
\usepackage{mathrsfs} 

\begin{document}
\begin{frontmatter}
\title{A non-cooperative Pareto-efficient solution to the one-shot Prisoner's Dilemma}
\author{Haoyang Wu\corauthref{cor}}
\corauth[cor]{Wan-Dou-Miao Research Lab, Suite 1002, 790 WuYi Road,
Shanghai, 200051, China.} \ead{hywch@mail.xjtu.edu.cn} \ead{Tel:
86-18621753457}

\begin{abstract}
The Prisoner's Dilemma is a simple model that captures the essential
contradiction between individual rationality and global rationality.
Although the one-shot Prisoner's Dilemma is usually viewed simple,
in this paper we will categorize it into five different types. For
the type-4 Prisoner's Dilemma game, we will propose a self-enforcing
algorithmic model to help non-cooperative agents obtain
Pareto-efficient payoffs. The algorithmic model is based on an
algorithm using complex numbers and can work in macro applications.
\end{abstract}

\end{frontmatter}

\section{Introduction}
The Prisoner's Dilemma (PD) is perhaps the most famous model in the
field of game theory. Roughly speaking, there are two sorts of PD:
one-shot PD and iterated PD. Nowadays a lot of studies on PD are
focused on the latter case. For example, Axelrod \cite{Axelrod1981}
investigated the evolution of cooperative behavior in well-mixed
populations of selfish agents by using PD as a paradigm. Nowak and
May \cite{Nowak1992} induced spatial structure in PD, i.e., agents
were restricted to interact with his immediate neighbors. Santos and
Pacheco \cite{Santos2005} found that when agents interacted
following scale-free networks, cooperation would become a dominating
trait throughout the entire range of parameters of PD. Perc and
Szolnoki \cite{Perc2008} proposed that social diversity could induce
cooperation as the dominating trait throughout the entire range of
parameters of PD.

Compared with the iterated PD, the one-shot PD is usually viewed
simple. In the original version of one-shot PD, two prisoners are
arrested by a policeman. Each prisoner must independently choose a
strategy between ``Confessing'' (denoted as strategy
``\emph{Defect}'') and ``Not confessing'' (denoted as strategy
``\emph{Cooperate}''). The payoff matrix of prisoners is shown in
Table 1. As long as two agents are rational, the unique Nash
equilibrium shall be (\emph{Defect}, \emph{Defect}), which results
in a Pareto-inefficient payoff $(P,P)$. That is the dilemma.

\emph{Table 1: The payoff matrix of PD, where }$T>R>P>S$, and
$R>(T+S)/2$. \emph{The first entry in the parenthesis denotes the
payoff of agent 1 and the second entry stands for the payoff of agent 2}.\\
\begin{tabular}{|c|c|c|}
\hline \backslashbox{agent 1}{agent 2} &
{\emph{Cooperate}}&{\emph{Defect}}
 \\\hline \emph{Cooperate} & (R, R) & (S, T)
\\ \emph{Defect} & (T, S) & (P, P)
\\ \hline
\end{tabular}

In 1999, Eisert \emph{et al} \cite{Eisert1999} proposed a quantum
model of one-shot PD (denoted as EWL model). The EWL model showed
``quantum advantages'' as a result of a novel quantum Nash
equilibrium, which help agents reach the Pareto-efficient payoff
$(R,R)$. Hence, the agents escape the dilemma. In 2002, Du \emph{et
al} \cite{Du2002} gave an experiment to carry out the EWL model.

So far, there are some criticisms on EWL model: 1) It is a new game
which has new rules and thus has no implications on the original
one-shot PD \cite{Enk2002}. 2) The quantum state serves as a binding
contract which let the players chooses one of the two possible moves
(\emph{Cooperate} or \emph{Defect}) of the original game. 3) In the
full three-parameter strategy space, there is no such quantum Nash
equilibrium \cite{BH2001} \cite{Flitney2007}.

Besides these criticisms, here we add another criticism: in the EWL
model, the arbitrator is required to perform quantum measurements to
readout the messages of agents. This requirement is unreasonable for
common macro disciplines such as politics and economics, because the
arbitrator should play a neutral role in the game: His reasonable
actions should only receive agents' strategies and assign payoffs to
agents. Put differently, if the arbitrator is willing to work with
an additional quantum equipment which helps agents to obtain the
Pareto-efficient payoffs $(R, R)$, then \emph{why does not he
directly assign the Pareto-efficient payoffs to the agents}?

Motivated by these criticisms, this paper aims to investigate
whether a Pareto-efficient outcome can be reached by non-cooperative
agents in macro applications. Note that a non-cooperative game is
one in which players make decisions independently. Thus, while they
may be able to cooperate, any cooperation must be self-enforcing
\cite{wiki}.

The rest of this paper is organized as follows: in Section 2 we will
propose an algorithmic model, where the arbitrator does not have to
work with some additional quantum equipment (Note: here we do not
aim to solve the first three criticisms on the EWL model, because
these criticisms are irrelevant to the algorithmic model). In
Section 3, we will categorize the one-shot PD into five different
types, and claim that the agents can self-enforcingly reach the
Pareto-efficient outcome for the case of type-4 PD by using the
algorithmic model. The Section 4 gives some discussions. The last
section draws conclusion.

\section{An algorithmic model}
As we have pointed out above, for macro applications, it is
unreasonable to require the arbitrator act with some additional
quantum equipment. In what follows, firstly we will amend the EWL
model such that the arbitrator works in the same way as he does in
classical environments, then we will propose an algorithmic version
of the amended EWL model.

\subsection{The amended EWL model}
Let the set of two agents be $N=\{1, 2\}$. Following formula (4) in
Ref. \cite{Flitney2007}, two-parameter quantum strategies are drawn
from the set:
\begin{equation*}
\hat{\omega}(\theta,\phi)\equiv \begin{bmatrix}
  e^{i\phi}\cos(\theta/2) & i\sin(\theta/2)\\
  i\sin(\theta/2) & e^{-i\phi}\cos(\theta/2)
\end{bmatrix},
\end{equation*}
$\hat{\Omega}\equiv\{\hat{\omega}(\theta,\phi):\theta\in[0,\pi],\phi\in[0,\pi/2]\}$,
$\hat{J}\equiv\cos(\gamma/2)\hat{I}\otimes
\hat{I}+i\sin(\gamma/2)\hat{\sigma_{x}}\otimes \hat{\sigma_{x}}$
(where $\gamma$ is an entanglement measure, $\hat{\sigma_{x}}$ is
the Pauli matrix, $\otimes$ is tensor product),
$\hat{I}\equiv\hat{\omega}(0,0)$,
$\hat{D}\equiv\hat{\omega}(\pi,\pi/2)$,
$\hat{C}\equiv\hat{\omega}(0,\pi/2)$.

Without loss of generality, we assume:\\
1) Each agent $j\in N$ has a quantum coin (qubit), a classical card
and a channel connected to the arbitrator. The basis vectors
$|C\rangle=[1,0]^{T}$, $|D\rangle=[0,1]^{T}$ of a quantum coin
denote head up and tail
up respectively.\\
2) Each agent $j\in N$ independently performs a local unitary
operation on his/her own quantum coin. The set of agent $j$'s
operation is $\hat{\Omega}_{j}=\hat{\Omega}$. A strategic operation
chosen by agent $j$ is denoted as
$\hat{\omega}_{j}\in\hat{\Omega}_{j}$. If
$\hat{\omega}_{j}=\hat{I}$, then
$\hat{\omega}_{j}(|C\rangle)=|C\rangle$,
$\hat{\omega}_{j}(|D\rangle)=|D\rangle$; If
$\hat{\omega}_{j}=\hat{D}$, then
$\hat{\omega}_{j}(|C\rangle)=|D\rangle$,
$\hat{\omega}_{j}(|D\rangle)=|C\rangle$. $\hat{I}$ denotes
``\emph{Not flip}'', $\hat{D}$ denotes ``\emph{Flip}''. \\
3) The two sides of a card are denoted as Side 0 and Side 1. The
messages written on the Side 0 (or Side 1) of card $j$ is denoted as
$card(j,0)$ (or $card(j,1)$). $card(j,0)$ represents
``\emph{Cooperate}'', and $card(j,1)$
represents ``\emph{Defect}''.\\
4) There is a device that can measure the state of two quantum coins
and send messages to the designer.

\begin{figure}[!t]
\centering
\includegraphics[height=2.4in,clip,keepaspectratio]{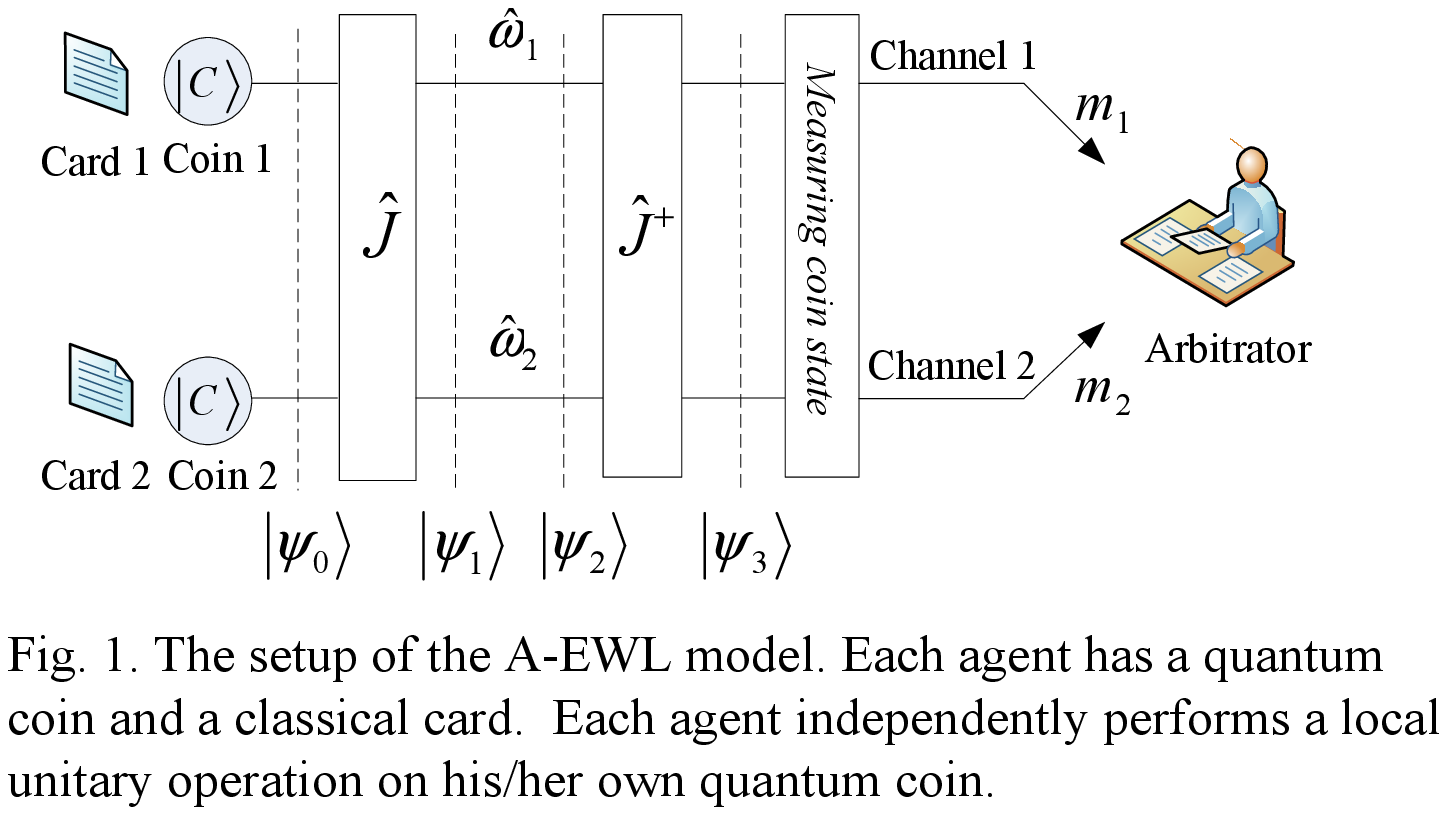}
\end{figure}

Fig. 1 shows the amended version of EWL model (denoted as the A-EWL
model). Its working steps are defined as follows: \\
Step 1: The state of each quantum coin is set as $|C\rangle$. The
initial state of the two quantum coins is
$|\psi_{0}\rangle=|CC\rangle$.\\
Step 2: Let the two quantum coins
be entangled by $\hat{J}$. $|\psi_{1}\rangle=\hat{J}|CC\rangle$.\\
Step 3: Each agent $j$ independently performs a local unitary
operation $\hat{\omega}_{j}$ on his own quantum coin.
$|\psi_{2}\rangle=[\hat{\omega}_{1}\otimes\hat{\omega}_{2}]\hat{J}|CC\rangle$.\\
Step 4: Let the two quantum coins be disentangled by $\hat{J}^{+}$.
$|\psi_{3}\rangle=\hat{J}^{+}[\hat{\omega}_{1}\otimes\hat{\omega}_{2}]
\hat{J}|CC\rangle$.\\
Step 5: The device measures the state of the two quantum coins and
sends $card(j,0)$ (or $card(j,1)$) as the message $m_{j}$ to the
arbitrator if the collapsed state of quantum coin $j$ is $|C\rangle$ (or $|D\rangle$).\\
Step 8: The arbitrator receives the overall message $m=(m_{1},
m_{2})$ and assigns payoffs to the two agents according to Table 1.
END.

In the A-EWL model, the assumed device performs quantum measurements
and sends messages to the arbitrator on behalf of agents. Thus, the
arbitrator needs not work with an additional quantum equipment as
EWL model requires, i.e., the arbitrator works in the same way as
before. It should be emphasized that the A-EWL model does not aim to
solve the criticisms on the EWL model as specified in the
Introduction. We propose the A-EWL model only for the following
simulation process, which is a key part of the algorithmic model.

Since quantum operations can be simulated classically by using
complex numbers, the A-EWL model can also be simulated. In what
follows we will give matrix representations of quantum states and
then propose an algorithmic version of A-EWL model.

\subsection{Matrix representations of quantum states}
In quantum mechanics, a quantum state can be described as a vector.
For a two-level system, there are two basis vectors: $[1,0]^{T}$ and
$[0,1]^{T}$. In the beginning, we define:
\begin{align*}
|CC\rangle=[1,0,0,0]^{T}, |CD\rangle=[0,1,0,0]^{T},
|DC\rangle=[0,0,1,0]^{T}, |DD\rangle=[0,0,0,1]^{T}.
\end{align*}
\begin{align*}
\hat{J}=\begin{bmatrix}
  \cos(\gamma/2) & 0 &  0 & i\sin(\gamma/2)\\
  0 & \cos(\gamma/2) & i\sin(\gamma/2) & 0 \\
  0 & i\sin(\gamma/2)& \cos(\gamma/2) &  0 \\
  i\sin(\gamma/2) & 0 & 0 & \cos(\gamma/2)
\end{bmatrix},
\;\gamma\in[0,\pi/2].
\end{align*}
For $\gamma=\pi/2$,
\begin{align*}
\hat{J}_{\pi/2}=\frac{1}{\sqrt{2}}\begin{bmatrix}
  1 & 0& 0& i\\
  0 & 1& i& 0\\
  0 & i& 1& 0\\
  i & 0& 0& 1
\end{bmatrix},
\hat{J}^{+}_{\pi/2}=\frac{1}{\sqrt{2}}\begin{bmatrix}
  1 & 0& 0& -i\\
  0 & 1& -i& 0\\
  0 & -i& 1& 0\\
  -i & 0& 0& 1
\end{bmatrix},
\end{align*}
where $\hat{J}^{+}_{\pi/2}$ is the conjugate of $\hat{J}_{\pi/2}$.

\textbf{Definition 1}:
$\psi_{1}\equiv\hat{J}|CC\rangle=\begin{bmatrix}
  \cos(\gamma/2)\\
  0\\
  0\\
  i\sin(\gamma/2)
\end{bmatrix}$.\\
Since only two values in $\psi_{1}$ are non-zero, we only need to
calculate the leftmost and rightmost column of
$\hat{\omega}_{1}\otimes\hat{\omega}_{2}$ to derive
$\psi_{2}=[\hat{\omega}_{1}\otimes\hat{\omega}_{2}]\psi_{1}$.

\textbf{Definition 2}: $\psi_{3}\equiv \hat{J}^{+}\psi_{2}$.

Suppose $\psi_{3}=[\eta_{1}, \cdots, \eta_{4}]^{T}$, let
$\Delta=[|\eta_{1}|^{2}, \cdots, |\eta_{4}|^{2}]$. It can be easily
checked that $\hat{J}$, $\hat{\omega}_{1}$, $\hat{\omega}_{2}$ and
$\hat{J}^{+}$ are all unitary matrices. Hence, $|\psi_{3}|^{2}=1$.
Thus, $\Delta$ can be viewed as a probability distribution over the
states $\{|CC\rangle, |CD\rangle, |DC\rangle, |DD\rangle\}$.

\subsection{An algorithmic model}
Based on the matrix representations of quantum states, here we will
propose an algorithmic model that simulates the A-EWL model. Since
the entanglement measurement $\gamma$ is a control factor, it can be
simply set as its maximum $\pi/2$. The input and output of the
algorithmic model are shown in Fig. 2. A \emph{Matlab} program is
shown in Fig. 3(a)-(d).

\begin{figure}[!t]
\centering
\includegraphics[height=1.7in,clip,keepaspectratio]{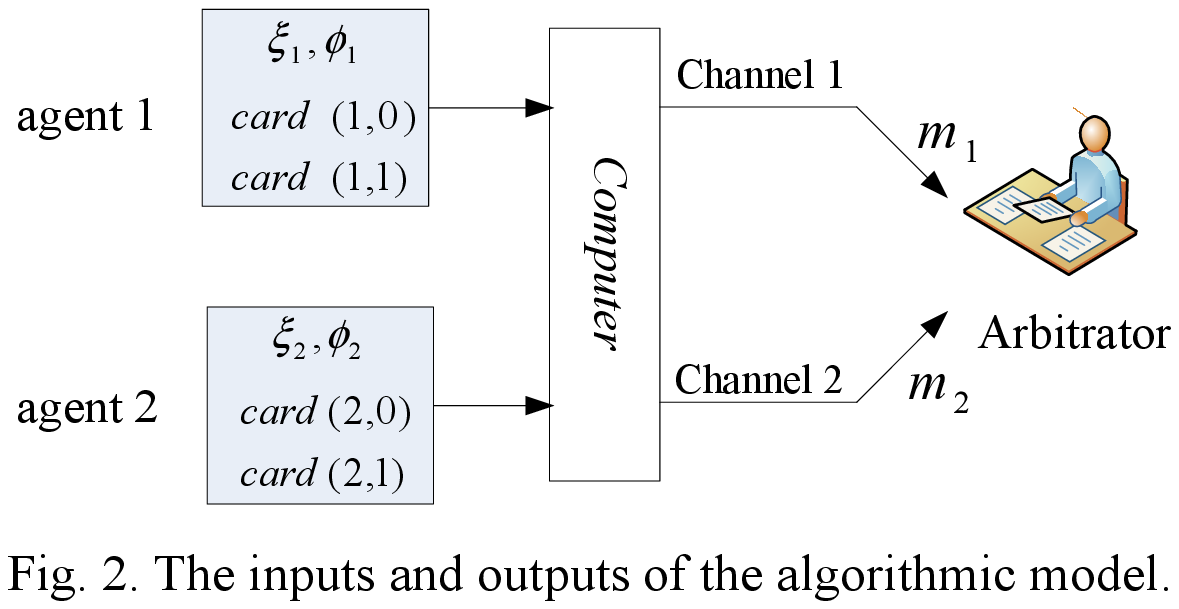}
\end{figure}

\textbf{Input}:\\
1) $\xi_{j}, \phi_{j}$, $j=1, 2$: the parameters of agent $j$'s
local operation $\hat{\omega}_{j}$,
$\xi_{j}\in[0,\pi],\phi_{j}\in[0,\pi/2]$.\\
2) $card(j,0),card(j,1)$, $j=1,2$: the messages written on the two
sides of agent $j$'s card. $card(j,0)$ and $card(j,1)$ represent
\emph{Cooperate} and \emph{Defect} respectively.

\textbf{Output}:\\
$m_{j}\in$ \{$card(j,0),card(j,1)$\}, $j=1, 2$: agent $j$'s message
that is sent to the arbitrator.

\textbf{Procedures of the algorithmic model}:\\
Step 1: Reading two parameters $\xi_{j}$ and $\phi_{j}$ from each
agent $j$ (See Fig. 3(a)).\\
Step 2: Computing the leftmost and rightmost columns of
$\hat{\omega}_{1}\otimes\hat{\omega}_{2}$ (See Fig. 3(b)).\\
Step 3: Computing
$\psi_{2}=[\hat{\omega}_{1}\otimes\hat{\omega}_{2}]\hat{J}_{\pi/2}|CC\rangle$,
$\psi_{3}=\hat{J}^{+}_{\pi/2}\psi_{2}$, and the probability
distribution $\Delta$ (See Fig. 3(c)).\\
Step 4: Randomly choosing a state from the set of all four possible
states $\{|CC\rangle, |CD\rangle, |DC\rangle, |DD\rangle\}$
according to the probability distribution $\Delta$.\\
Step 5: For each $j\in I$, the computer sends $card(j,0)$ (or
$card(j,1)$) as message $m_{j}$ to the arbitrator through channel
$j$ if the $j$-th element of the chosen state is $|C\rangle$ (or
$|D\rangle$) (See Fig. 3(d)).

\section{Five types of one-shot PD}
Since its beginning, PD has been generalized to many disciplines
such as politics, economics, sociology, biology and so on. Despite
these widespread applications, people seldom care how the payoffs of
agents are determined. For example, Axelrod \cite{Axelrod1981} used
the word ``yield'' to describe how the agents obtained the payoffs.
Nowak and May \cite{Nowak1992} used the word ``get'', and Santos and
Pacheco \cite{Santos2005} used the word ``receive'' respectively.

One may think that such question looks trivial at first sight.
However, as we will show in this section, there exists an
interesting story behind this question. In what follows, we will
categorize the one-shot PD into five different types.

\textbf{Type-1 PD}:\\
1) There are two agents and no arbitrator in the game. \\
2) The strategies of agents are actions performed by agents. The
agents' payoffs are determined by the outcomes of these actions and
satisfy Table 1.

For example, let us neglect the United Nation and consider two
countries (e.g., US and Russia) confronted the problem of nuclear
disarmament. The strategy \emph{Cooperate} means ``Obeying
disarmament'', and \emph{Defect} means ``Refusing disarmament''. If
the payoff matrix confronted by the two countries satisfies Table 1,
the nuclear disarmament game is a type-1 PD.

\textbf{Type-2 PD}:\\
1) There are two agents and an arbitrator in the game.\\
2) The strategies of agents are actions performed by agents. The
arbitrator observes the outcomes of actions and assign payoffs to
the agents according to Table 1.

For example, let us consider a taxi game. Suppose there are two taxi
drivers and a manager. Two drivers drive a car in turn, one in day
and the other in night. The car's status will be very good, ok or
common if the number of drivers who maintain the car is two, one or
zero respectively. The manager observes the car's status and assigns
rewards $R_{2}$, $R_{1}$, $R_{0}$ to each driver respectively, where
$R_{2}>R_{1}>R_{0}$. The whole cost of maintenance is $c$. Let the
strategy \emph{Cooperate} denote ``Maintain'', and \emph{Defect}
denote ``Not maintain''. The payoff matrix can be represented as
Table 2. If Table 2 satisfies the conditions in Table 1, the taxi
game is a type-2 PD.

\emph{Table 2: The payoff matrix of type-2 PD}.\\
\begin{tabular}{|c|c|c|}
\hline \backslashbox{agent 1}{agent 2} &
{\emph{Cooperate}}&{\emph{Defect}}
 \\\hline \emph{Cooperate} & ($R_{2}-c/2, R_{2}-c/2$) & ($R_{1}-c, R_{1}$)
\\ \emph{Defect} & ($R_{1}, R_{1}-c$) & ($R_{0}, R_{0}$)
\\ \hline
\end{tabular}

\textbf{Type-3 PD}:\\
1) There are two agents and an arbitrator in the game.\\
2) The strategy of each agent is not an action, but a message that
can be sent to the arbitrator through a channel. The arbitrator
receives two messages and assign payoffs to the agents according
to Table 1.\\
3) Two agents cannot communicate with each other.

For example, suppose two agents are arrested separately and required
to report their crime information to the arbitrator through two
channels independently. If the arbitrator assigns payoffs to agents
according to Table 1, this game is a type-3 PD.

\textbf{Type-4 PD}:\\
Conditions 1-2 are the same as those in type-3 PD. \\
3) Two agents can communicate with each other. \\
4) Before sending messages to the arbitrator, two agents can
construct the algorithmic model specified in Fig. 2. Each agent $j$
can observe whether the other agent participates the algorithmic
model or not: whenever the other agent takes back his channel, agent
$j$ will do so and sends his message $m_{j}$ to the arbitrator
directly.

\emph{Remark 1}: At first sight, the conditions of type-4 PD is
complicated. However, these conditions are not restrictive when the
arbitrator communicate with agents indirectly and cannot separate
them. For example, suppose the arbitrator and agents are connected
by Internet, then all conditions of type-4 PD can be satisfied in
principle.

The type-4 PD works in the following way:\\
\emph{Stage 1: (Actions of two agents)} For each agent $j\in N$,
he faces two strategies: \\
$\bullet$ $S(j,0)$: Participate the algorithmic model, i.e., leave
his channel to the computer, and submit $\xi_{j}, \phi_{j},
card(j,0), card(j,1)$ to the computer;\\
$\bullet$ $S(j,1)$: Not participate the algorithmic model, i.e.,
take back his channel, and submit $m_{j}$ to the arbitrator directly.\\
According to condition 4, the algorithmic model is triggered if and
only if both two agents participate it.\\
\emph{Stage 2: (Actions of the arbitrator)} The arbitrator receives
two messages and assigns payoffs to agents according to Table 1.

In type-4 PD, from the viewpoints of the arbitrator, he acts in the
same way as before, i.e., nothing is changed. However, the payoff
matrix confronted by two agents is now changed to Table 3. For each
entry of Table 3, we give the corresponding explanation as follows:

\emph{Table 3: The payoff matrix of two agents by constructing the
algorithmic model, where} $R,P$ \emph{are defined in Table 1}, $R>P$.\\
\begin{tabular}{|c|c|c|}
\hline \backslashbox{agent 1}{agent 2} & {$S(2,0)$}&{$S(2,1)$}
 \\\hline $S(1,0)$ & (R, R) & (P, P)
\\ $S(1,1)$ & (P, P) & (P, P)
\\ \hline
\end{tabular}

1) $(S(1,0),S(2,0))$: This strategy profile means two agents both
participate the algorithmic model and submit parameters to the
computer. According to Ref. \cite{Eisert1999}, for each agent $j\in
N$, his dominant parameters are $\xi_{j}=0$
and $\phi_{j}=\pi/2$, which result in a Pareto-efficient payoff $(R,R)$. \\
2) $(S(1,0),S(2,1))$: This strategy profile means agent 1
participates the algorithmic model, but agent 2 takes back his
channel and submits a message to the arbitrator directly. Since
agent 1 can observe agent 2's action, in the end, both agents will
take back their channels and submit messages to the arbitrator
directly. Obviously, the dominant message of each agent $j$ is
$card(j,1)$, and the arbitrator will assign the Pareto-inefficient
payoff $(P,P)$ to agents. \\
3) $(S(1,1),S(2,0))$: This strategy profile is similar to the above
case. The arbitrator will assign $(P, P)$ to two agents.\\
4) $(S(1,1),S(2,1))$: This strategy profile means two agents both
take back their channels and send messages to the arbitrator
directly. This case is similar to the case 2. The arbitrator will
assign $(P, P)$ to two agents.

From Table 3, it can be seen that $(S(1,0),S(2,0))$ and
$(S(1,1),S(2,1))$ are two Nash equilibria, and the former is
Pareto-efficient. As specified by Telser (Page 28, Line 2,
\cite{Telser1980}), ``\emph{A party to a self-enforcing agreement
calculates whether his gain from violating the agreement is greater
or less than the loss of future net benefits that he would incur as
a result of detection of his violation and the consequent
termination of the agreement by the other party.}'' Since two
channels have been controlled by the computer in Stage 1, in the end
$(S(1,0),S(2,0))$ is a self-enforcing Nash equilibrium and the
Pareto-efficient payoff $(R,R)$ is the unique Nash equilibrium
outcome. In this sense, the two agents escape the dilemma.

\textbf{Type-5 PD}:\\
Conditions 1-3 are the same as those in type-4 PD. \\
4) The last condition of type-4 PD does not hold.\\
For this case, although the two agents can communicate before moving
and agree that collaboration is good for each agent, they will
definitely choose (\emph{Defect}, \emph{Defect}) as if they are
separated. Thus, the agents cannot escape the dilemma.

\section{Discussions}
The algorithmic model revises common understanding on the one-shot
PD. Here we will discuss some possible doubts about it.

\emph{Q1}: The type-4 PD seems to be a cooperative game because in
condition 4, the algorithmic model constructed by two agents acts as
a correlation between agents.

\emph{A1}: From the viewpoints of agents, the game is different from
the original one-shot PD, since the payoff matrix confronted by the
two agents has been changed from Table 1 to Table 3. But from the
viewpoints of the arbitrator, nothing is changed. Thus, the
so-called correlation between two agents is indeed
\emph{unobservable} to the arbitrator. Put differently, the
arbitrator cannot prevent agents from constructing the algorithmic model.\\
On the other hand, since each agent can freely choose not to
participate the algorithmic model and send a message to the
arbitrator directly in Stage 1, the algorithmic model is
self-enforcing and thus still a non-cooperative game.

\emph{Q2}: After the algorithmic model is triggered, can it simply
send $(card(1,0)$, $card(2,0))$ to the arbitrator instead of running
Steps 1-5?

\emph{A2}: The algorithmic model enlarges each agent's strategy
space from the original strategy space \emph{\{Cooperate, Defect\}}
to a two-dimensional strategy space $[0,\pi]\times[0,\pi/2]$, and
generates the Pareto-efficient payoff $(R,R)$ in Nash equilibrium.
The enlarged strategy space includes the original strategy space of
one-shot PD: the strategy (\emph{Cooperate, Cooperate}),
(\emph{Cooperate, Defect}), (\emph{Defect, Cooperate}),
(\emph{Defect, Defect}) in the original PD correspond to the
strategy $((0, 0), (0, 0))$, $((0, 0), (\pi, \pi/2))$, $((\pi,
\pi/2), (0, 0))$, $((\pi, \pi/2), (\pi, \pi/2))$ in the algorithmic
model respectively, since $\hat{I}=\hat{\omega}(0,0)$, $\hat{D}=
\hat{\omega}(\pi,\pi/2)$.\\
However, the idea in this question restricts each agent's strategy
space from the original strategy space \emph{\{Cooperate, Defect\}}
to a single strategy \emph{Cooperate}. In this sense, two agents are
required to sign a binding contract to do so. This is beyond the
range of non-cooperative game.

\emph{Remark 2}: The algorithmic model is not suitable for type-1
and type-2 PD, because the computer cannot perform actions on behalf
of agents. The algorithmic model is not applicable for type-3 PD
either because two agents are separated, thereby the algorithmic
model cannot be constructed. For the case of type-5 PD, the
algorithmic model is not applicable because condition 4 in type-4 PD
is vital and indispensable.

\section{Conclusion}
In this paper, we categorize the well-known one-shot PD into five
types and propose an algorithmic model to help two non-cooperative
agents self-enforcingly escape a special type of PD, i.e., the
type-4 PD. The type-4 PD is justified when the arbitrator
communicate with the agents indirectly through some channels, and
each agent's strategy is not an action, but a message that can be
sent to the arbitrator. With the rapid development of Internet, more
and more type-4 PD games will be seen.

One point is important for the novel result: Usually people may
think the two payoff matrices confronted by agents and the
arbitrator are the same (i.e., Table 1). However we argue that for
the case of type-4 PD, the two payoff matrices can be different: The
arbitrator still faces Table 1, but the agents can self-enforcingly
change their payoff matrix to Table 3 by virtue of the algorithmic
model, which leads to a Pareto-efficient payoff.

\section*{Acknowledgments}

The author is very grateful to Ms. Fang Chen, Hanyue Wu
(\emph{Apple}), Hanxing Wu (\emph{Lily}) and Hanchen Wu
(\emph{Cindy}) for their great support.


\newpage
\begin{figure}[!t]
\centering
\includegraphics[height=2.6in,clip,keepaspectratio]{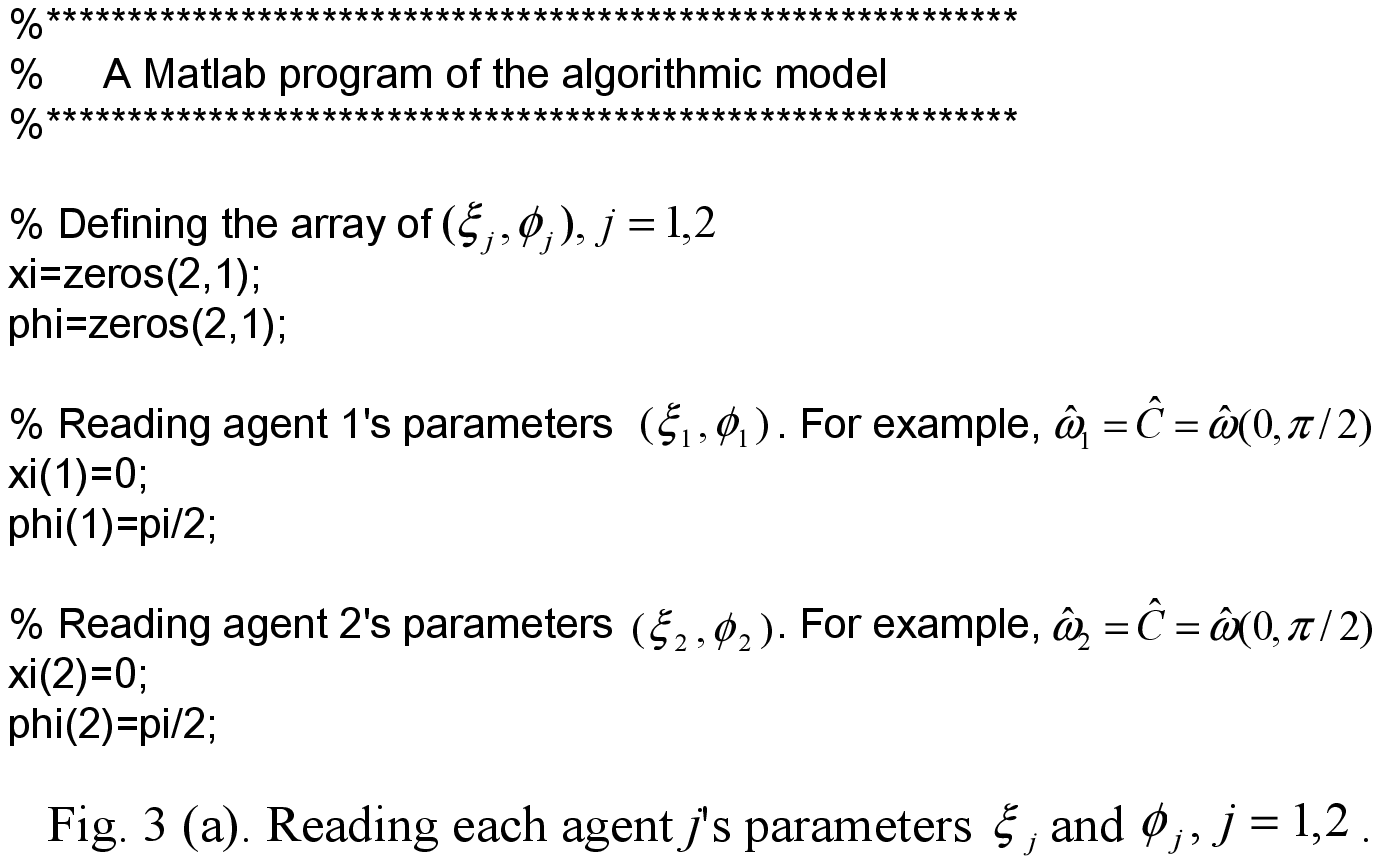}
\end{figure}

\begin{figure}[!t]
\centering
\includegraphics[height=4.5in,clip,keepaspectratio]{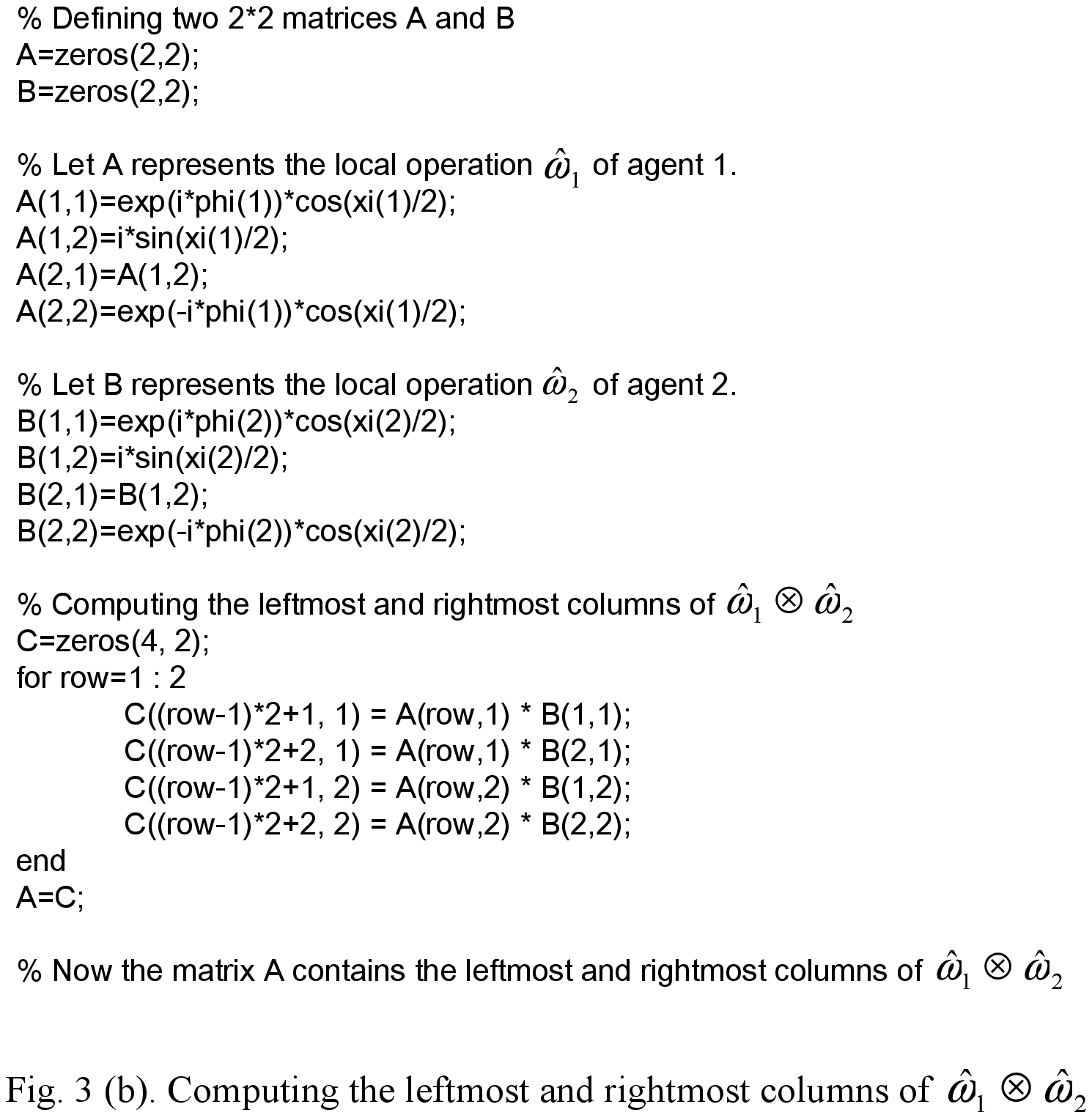}
\end{figure}

\newpage
\begin{figure}[!t]
\centering
\includegraphics[height=2.4in,clip,keepaspectratio]{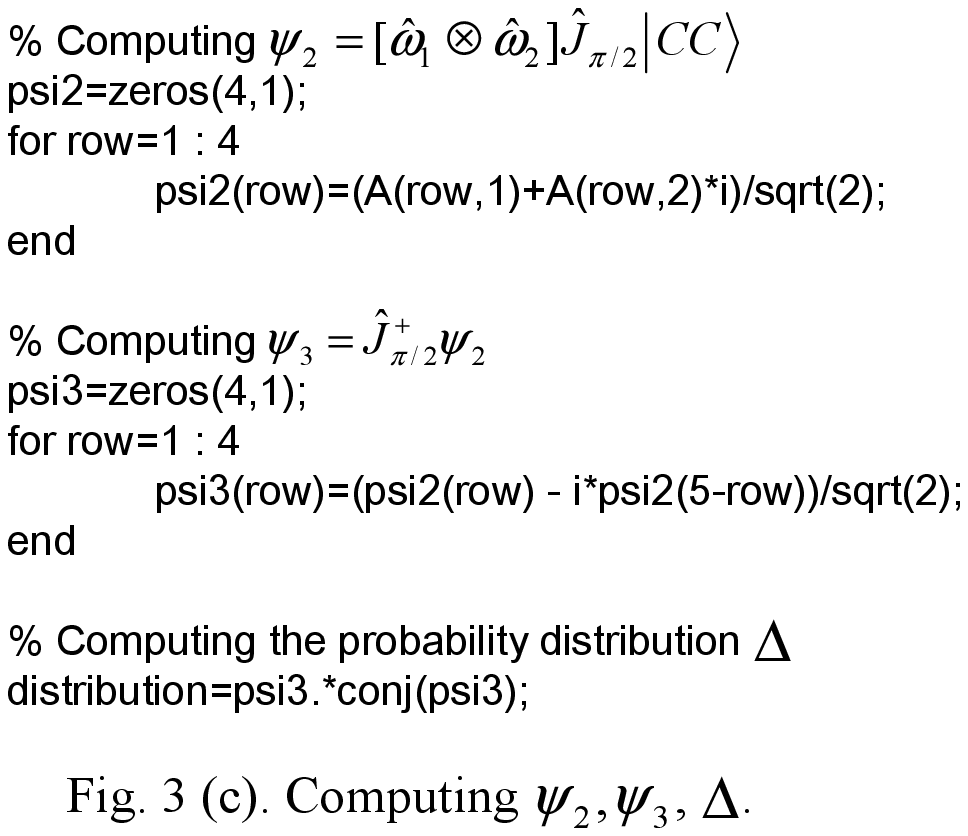}
\end{figure}

\begin{figure}[!t]
\centering
\includegraphics[height=5.5in,clip,keepaspectratio]{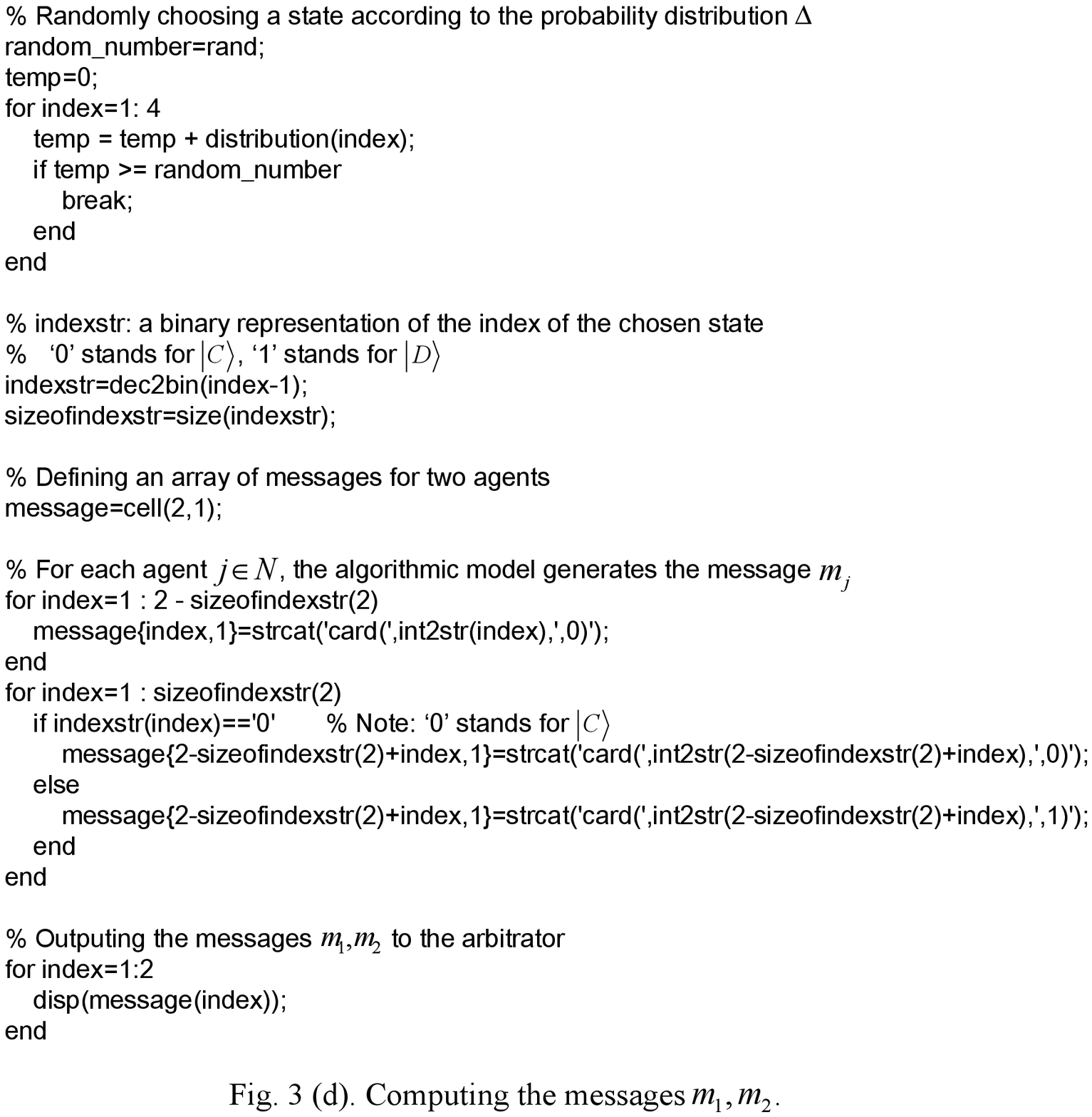}
\end{figure}

\end{document}